\documentclass[preprint]{elsarticle}
\usepackage{geometry}                
\usepackage[parfill]{parskip}    
\usepackage{graphicx}
\usepackage{lineno}
\usepackage{epstopdf}
\usepackage{xspace}
\usepackage{amsmath}
\newcommand{\mm}{\ensuremath{\rm \,mm}\xspace} 
\newcommand{\um}{{\ensuremath{\,\mu\rm m}\xspace} }  
\newcommand{\nm}{\ensuremath{\rm \,nm}\xspace}

\newcommand{\ms}{\ensuremath{{\rm \,ms}}\xspace} 
\newcommand{\ns}{\ensuremath{{\rm \,ns}}\xspace}

\newcommand{\Hz}{\ensuremath{{\rm \,Hz}}\xspace} 
\newcommand{\kHz}{\ensuremath{{\rm \,kHz}}\xspace}
\newcommand{\MHz}{\ensuremath{{\rm \,MHz}}\xspace}
\newcommand{\GHz}{\ensuremath{{\rm \,GHz}}\xspace}

\newcommand{\gevc }{\ensuremath{{\mathrm{\,GeV\!/}c}}\xspace}  

\title{The Timepix Telescope for High Performance Particle Tracking}

\author[rio]{K.~Akiba}
\author[cern]{P.~Ronning}
\author[nikhef]{M.~van~Beuzekom}
\author[nikhef]{V.~van~Beveren}
\author[manchester]{S.~Borghi}
\author[nikhef]{H.~Boterenbrood}
\author[cern]{J.~Buytaert}
\author[cern]{P.~Collins} 
\author[santiago]{A.~Dosil~Su\'arez}
\author[cern]{R.~Dumps}
\author[glasgow]{L.~Eklund}
\author[santiago]{D.~Esperante}
\author[santiago]{A.~Gallas}
\author[oxford]{H.~Gordon}
\author[nikhef]{B.~van~der~Heijden}
\author[manchester]{C.~Hombach}
\author[glasgow]{D.~Hynds}
\author[oxford]{M.~John}
\author[moscow]{A.~Leflat}
\author[oxford]{Y.~Li}
\author[glasgow]{I.~Longstaff}
\author[glasgow]{A.~Morton}
\author[cern]{N.~Nakatsuka}
\author[oxford]{A.~Nomerotski}
\author[manchester]{C.~Parkes}
\author[santiago]{E.~Perez~Trigo}
\author[dls]{R.~Plackett\corref{cor1}} 
\author[warwick]{M.~M.~Reid}
\author[santiago]{P.~Rodriguez~Perez}
\author[cern]{H.~Schindler}
\author[krakow]{T.~Szumlak}
\author[nikhef]{P.~Tsopelas}
\author[santiago]{C.~V\'azquez~Sierra}
\author[bristol]{J.~Velthuis}
\author[krakow]{M.~Wysoki\'nski}

\address[rio]{Federal University of Rio de Janeiro, Rio de Janeiro, Brazil}
\address[cern]{CERN, the European Organisation for Nuclear Research, Geneva, Switzerland}
\address[nikhef]{NIKHEF, Amsterdam, Netherlands}
\address[manchester]{University of Manchester, Manchester, Lancashire, UK}
\address[santiago]{Universidade de Santiago de Compostela, Santiago de Compostela, Spain}
\address[oxford]{University of Oxford, Oxfordshire, UK}
\address[glasgow]{Glasgow University, Glasgow, Lanarkshire, UK}
\address[moscow]{Lomonosov Moscow State University, Moscow, Russia}
\address[dls]{Diamond Light Source Ltd. Didcot, Oxfordshire, UK}
\address[warwick]{University of Warwick, Coventry, UK}
\address[krakow]{AGH University of Science and Technology, Krakow, Poland}
\address[bristol]{ Bristol University, Bristol, Avon, UK}

\cortext[cor1]{Corresponding author\\  Email: richard.plackett@cern.ch\\ Telephone: +44 1235 778163}

\begin{document}

\modulolinenumbers[5]		
\linenumbers				

\begin{abstract}

The Timepix particle tracking telescope has been developed as part of the LHCb VELO Upgrade project, supported by the Medipix Collaboration and the AIDA framework.  It is a primary piece of infrastructure for the VELO Upgrade project and is being used for the development of new sensors and front end technologies for several upcoming LHC trackers and vertexing systems.  The telescope is designed around the dual capability of the Timepix ASICs to provide information about either the deposited charge or the timing information from tracks traversing the $14 \times 14 \mm$ matrix of $55 \times 55\um$ pixels.   The rate of reconstructed tracks available is optimised by taking advantage of the shutter driver readout architecture of the Timepix chip, operated with existing readout systems.  Results of tests conducted in the SPS North Area beam facility at CERN show that the telescope typically provides reconstructed track rates during the beam spills of between 3.5 and 7.5\kHz, depending on beam conditions.  The tracks are time stamped with 1\ns resolution with an efficiency of above 98\% and provide a pointing resolution at the centre of the telescope of $\sim$1.6\um.  By dropping the time stamping requirement the rate can be increased to $\sim$15\kHz, at the expense of a small increase in background. The telescope infrastructure provides CO$_2$ cooling and a flexible mechanical interface to the device under test, and has been used for a wide range of measurements during the 2011-2012 data taking campaigns.

\end{abstract}

\begin{keyword}

Particle Tracking, Timepix, LHCb, VELO, Silicon Pixel Sensors, Vertexing, Medipix

\end{keyword}

\maketitle


\section{Introduction}
\label{introduction}

This paper reports on the design and performance of the Timepix Telescope, developed as part of the LHCb VELO Upgrade\cite{velo_upgrade} project with the support of the Medipix Collaboration\footnote{The Medipix Collaboration, http://medipix.web.cern.ch} and the AIDA\footnote{ Advanced European Infrastructures for Detectors at Accelerators, http://aida.web.cern.ch} research and development framework. The apparatus supports a continuing detector research program to develop detectors for future Particle Physics experiments, with a particular emphasis on the requirements of the LHCb VELO. To this end the telescope is designed to provide a high rate of tracks which are spatially precise and time resolved.  The track pointing resolution at the centre of the telescope is below 2\um and tracks are time stamped with a resolution of $~1\ns$.    The system is configured to allow easy integration of external readout systems with minimal adaptations to enable quick and flexible testing of a range of devices with widely varying specifications. There is a particular focus on devices with LHC style 40\MHz triggering requirements.   The track reconstruction rate, depending on the beam conditions and user requirements, has been measured at the SPS H8 beamline to be between 3.5 and 15.5\kHz during spills, allowing of the order of a million tracks to be accumulated at the device under test (DUT) during 5 to 25 minutes of data taking operation, depending on running conditions.

This paper presents a description of the system and sample results demonstrating its performance in terms of spatial and timing resolution, efficiency and particle flux.

\section{Description of Telescope Systems}

The dual functionality of the Timepix\cite{timepix} chip allows a single interface to be used to record either time or position tracking information from the pixel detector planes.  As the Timepix system can only operate with a limited clock frequency, a secondary timing system linked to coincident scintillators is used to achieve the desired time resolution and provide triggers for the DUTs.

\subsection{Core Tracking Elements}

\begin{figure}[h]
\centering
\includegraphics[width=14cm]{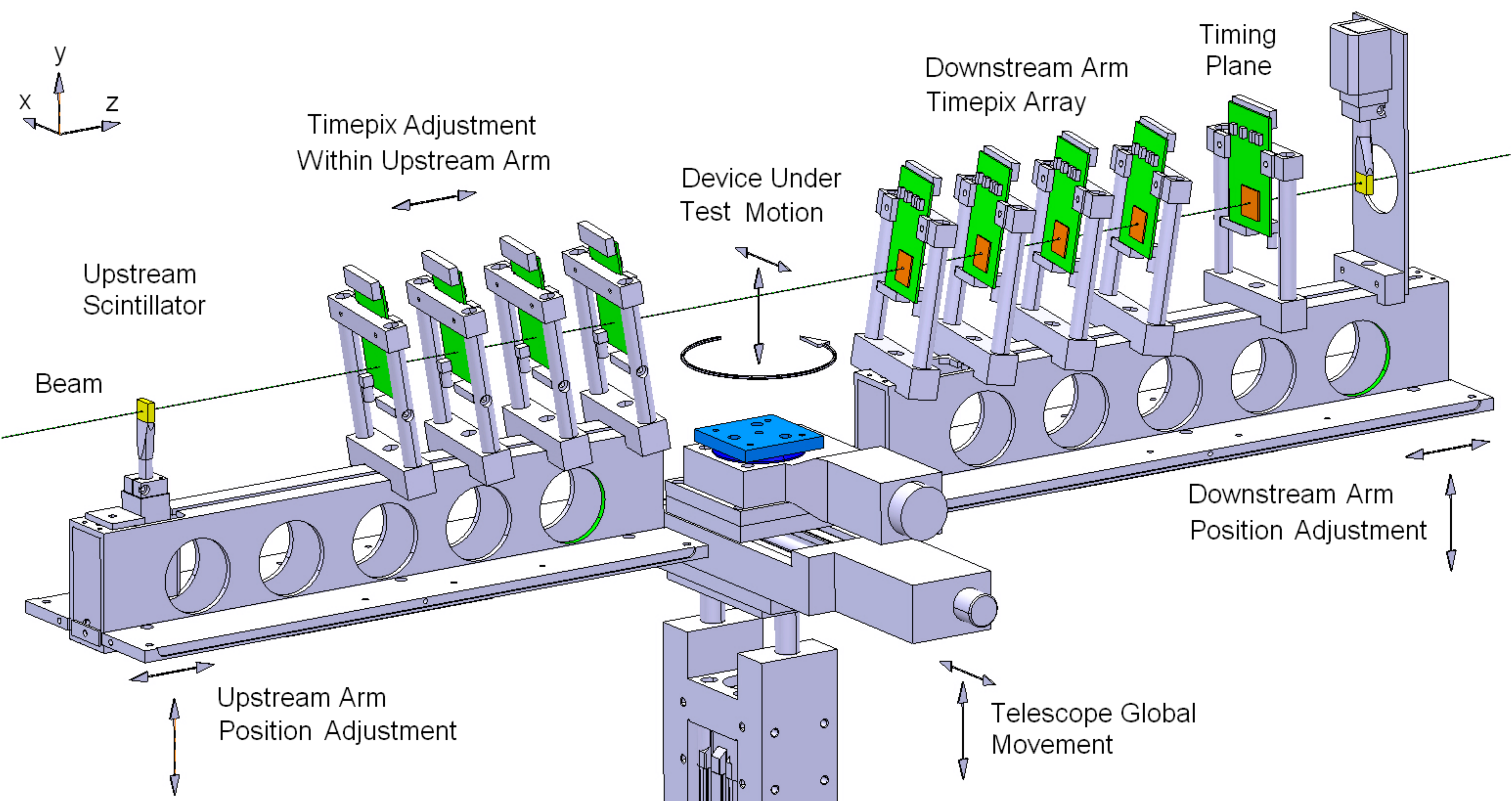}
\caption{Layout of the Timepix Telescope mechanics, pixel planes and scintillators with respect to the beam axis.}
\label{fig:telescope_schematic}
\end{figure}

The core of the telescope comprises eight Timepix silicon hybrid detectors arranged into two `arms' of four about a central DUT station together with a ninth plane which provides timing information.  The telescope makes maximum use of existing systems and software and is designed to be as flexible and automated as possible. The system exploits the Timepix chip's unique combination of small (55\um\!) square pixels and its ability to record either the charge deposited (in time over threshold or ToT mode) or the particle's time of arrival (in ToA mode) from the pixels.  A sketch of the complete telescope geometry is shown in Figure~\ref{fig:telescope_schematic}.  The eight pixel planes are angled to nine degrees in both horizontal and vertical axes to optimise the spatial resolution that can be obtained from the system.  This angle, given the pixel size and the 300\um sensor thickness, results in typical cluster sizes of three pixels, allowing the ToT information to be used to calculate a centre of gravity for the cluster and provide a reconstructed position with sub-pixel resolution\cite{telescope2009}.  The ninth plane recording the ToA information is mounted perpendicularly to the beam to reduce the charge sharing and provide a more prompt response from the pixel's discriminator. This plane is mounted downstream of the eight position tracking planes to avoid introducing material that limits the spatial resolution of the system due to multiple scattering.

The Timepix planes operate with an untriggered camera-like readout scheme, being continuously sensitive when a `shutter' signal is active and insensitive whilst the data is read off the pixel matrices. The Timepix devices are read by a series of RELAXD\cite{relaxd} readout systems, which are linked to a single DAQ PC via dedicated UDP Gigabit Ethernet links.  The four detector planes in each telescope arm are read in parallel by separate RELAXD systems, which are synchronised by an externally generated shutter.  In this configuration, the time required to read the data from the four pixel matrices is about 17\ms, allowing a maximum frame rate of 60\Hz. The shutter signal, which determines the active period of the telescope, is produced by a NIM logic system, described in detail in the next section.  The system automatically adjusts the time for which the shutter is open to compensate for variations in the beam flux keeping the number of tracks per shutter approximately constant and the occupancy of the detectors to a pre-defined level.   It is also possible to operate the telescope in a mode where the shutter times are kept fixed, which is of interest for certain DUT requirements and low beam intensities.

\subsection {Timing and Test Device Integration Systems }

\begin{figure}[h]
\centering
\includegraphics[width=12cm]{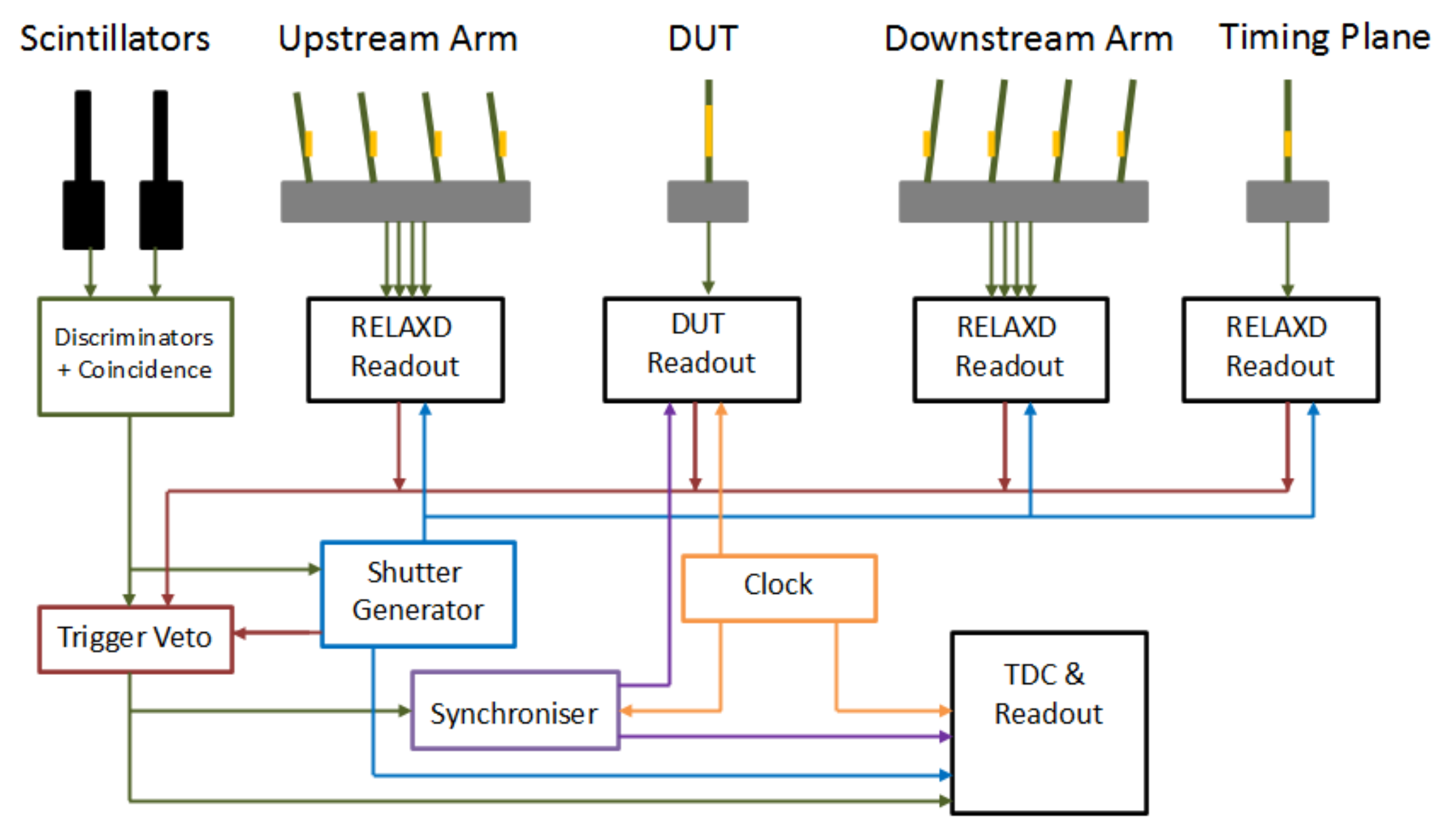}
\caption{The telescope timing system used, highlighting the principal components, namely the Timepix shutter, the readout systems' vetos, the raw triggers from the scintillators, the synchronised triggers and the system clock.}
\label{fig:timing_system}
\end{figure}

The telescope system has been designed to characterise detectors and sensors running either with a Medipix camera-like readout, or pipelined and triggered HEP style readout. The initial devices tested of the latter type were prototype LHCb strip sensors with Beetle\cite{beetle} chip readout and ATLAS FE-I4\cite{FEI4} hybrid pixel assemblies. Communication between the DUTs and the telescope is kept to a minimum, in order to ease the integration of the two systems. The DUT systems are provided with a trigger generated by the telescope's coincident scintillators mounted on the arms at the extreme ends of the telescope. The external readout can provide a busy signal to inhibit further triggers being generated if necessary.  The coincident signals forming the trigger can be synchronised to an arbitrary clock signal to provide a synchronised trigger as required by LHC devices.  The synchronised and unsynchronised triggers and the shutter open and close signals are recorded by a high precision Time to Digital Converter (TDC)\footnote{Module V1190N from CAEN Electronic Instrumentation} to allow the data from the tracking system to be effectively combined with the data from the external system in the analysis step.  This TDC data also allows the accurate times from the scintillator system ($\sim 1$\ns) to be allocated to the less precise time bins of particular Timepix tracks.  A detailed conceptual layout for the timing system is shown in Figure~\ref{fig:timing_system}.

As systems designed to operate in a pulsed system like the LHC often have inefficiencies when operating out of phase, this system allows this effect to be measured in an asynchronous beam such as that provided by CERN's SPS North Area.  The particles arrive with an essentially random phase with respect to the system clock and the TDC data can be used to produce a mapping of phase against efficiency, for example allowing sensitive measurements of a detector's front-end peaking times.  The fast timing and continuous sensitivity available to the telescope also enables a measurement of hits in a way that simulates consecutive LHC bunch crossings, allowing detailed measurements of the effect of a system's dead time and pile up.

Over the course of measurements performed in 2011 \& 2012 a large number of different systems and devices have been integrated with the telescope and tested.  These include Timepix assemblies with novel guard ring structures and tests to evaluate power-pulsing, irradiated and unirradiated Medipix3 assemblies, strip sensors with a Beetle readout, scintillating fibre modules, an LHCb TORCH prototype, ATLAS FE-I3 and FE-I4 assemblies and an HV-MAPS project.  This range of DUTs is possible due to the minimal and flexible integration made possible by the telescope's architecture.

\subsection{ Mechanics and Services}

The right hand coordinate system used to describe the telescope sets $z$ along the direction of the beam, $x$ horizontally and $y$ vertically.  Each of the telescope arms contains four angled planes, which have a small manual $z$ adjustment along a rail inside the arms.  The arms themselves can be adjusted and locked in $y$ and $z$.  The entire telescope structure can be moved in $x$ and $y$ remotely to allow the beam position to be determined quickly, or for the entire system to be removed from the beam without the need to enter the beam area.

The DUT stage is mounted between the two arms, although other DUTs can be mounted upstream or downstream as desired.  The stage provides micron accurate $x$ and $y$ translation and hundredth of a degree accurate rotation stages.  The DUT enclosure can weigh up to 50~kg and still be lifted by the $y$ stage, and can have a maximum $z$ extent of about 45~cm. The DUT enclosure is provided with two-phase CO$_2$\cite{co2} and Peltier cooling facilities capable of reaching $-20~^o$C and a dry nitrogen supply.  The cooling system has been developed in partnership with the telescope project and is capable of holding the DUT at a desired temperature with an accuracy of better than 1~$^o$C allowing irradiated DUTs to be tested.  The motion of the driven stages, the temperature, low voltage and sensor bias of the various systems is recorded and logged by an application running on a dedicated control PC to allow the state of the system to be recovered during the analysis.

\section{Software}

The analysis framework is designed to treat both the data and the algorithms in an object-oriented manner.
This means data is packaged into discrete blocks which are passed through a succession of task-specific algorithms.
To allow efficient processing of the raw data, the framework has two principal steps. 
The first step performs the low-level event-building and persists the formatted data into a bespoke class structure based on the ROOT framework\footnote{The ROOT Framework, http://root.cern.ch/}. The second step is the analysis which reads in the formatted data and performs the clustering and track-finding described in the next section.\\

Data from the telescope, the TDC, and the DUT are transferred from the various data acquisition disks of each subsystem to a central analysis PC.
The {\it event builder} amalgamates these data into a well defined data-block that contains all information for a given period of time.
The data is naturally discretised into the {\it shutters} during which the Timepix chips of the telescope are active.
In this way the event-building of the telescope data, for which the RELAXD system provides a sychronous readout, is straightforward.
The TDC records shutter opening and closing, as well as the passage of each charged particle in the scintillators with nanosecond precision.
The block of TDC time stamps between each shutter-opening and shutter-closing signals is associated with each telescope {\it event} (shutter) in a linear order.\\

The DUTs are of two general types, those which operate with a shutter mode in a similar manner to the telescope itself, and those which have continuous pipelined readout.
The first type, which include Timepix chips and irradiated Medipix3 devices are relatively simple to time-align with the telescope.
Devices that operate with pipelined continuous readout (e.g. devices using FE-I4 or Beetle readout) are more challenging as they write out data every 25\ns regardless of the state of the telescope shutters.
In this case, the event-building relies upon matching the TDC time stamps (which includes the shutter opening/closing information) to the stream of 25\ns-spaced events from the DUTs.

\section{Track Reconstruction}

To identify particle tracks from the shutter based Timepix data requires the transformation of information from the planes containing multiple pixel hits into a set of vectors in 3D space that can be projected onto the DUT.  This is achieved in three stages described below: clustering, pattern recognition and track fitting.

\subsection{Clustering}

The first step of the reconstruction sequence is to form clusters from groups of neighbouring pixels that register hits.  The clusters formed also take account of the ToT values, which are corrected with a surrogate function obtained from a charge calibration with test-pulses as described in~\cite{telescope2009}. This calibration allows the recovery of a linear relationship between the ToT and the deposited charge.  Typically the most probable value of the Landau distribution from the charge deposited by a minimum ionising particle is approximately 125 ADC counts. Clusters with values that deviate strongly from this are excluded in order to suppress high energy deposits from hadronic interactions in the sensor, delta rays and to remove the effects of noisy pixels.  
The final position of a multi-pixel cluster is calculated as the centre-of-gravity (CoG) of the pixel coordinates, using the ToT from the individual pixels as weighting factors.  The clustering algorithm is configurable, and the parameters used for the standard telescope running conditions are given in Table~\ref{tab:params}.

\subsection{Pattern Recognition}

The pattern recognition provides a method of selecting clusters to form tracks which is not computationally prohibitive.  The algorithm takes multiple scattering in the material of the telescope into account and passes the selected clusters onto the track fitter to be evaluated.

The approach taken for the pattern recognition is based on a k-Nearest Neighbours (k-NN) algorithm\cite{KDTree}, a method that classifies objects based on distances between space vectors. It is a data structure and search algorithm to find near neighbours (within a fixed radius) of a user supplied reference vector, from a fixed database of vectors in d-dimensional Euclidean space. On average it will take $\cal{O} ( \rm log(N))$ bisections to locate nearest neighbours, a significant reduction on the computational effort required to exhaustively search all N space points, needing a computational time proportional to $\cal{O} (\rm{N})$. We use the k-NN in our pattern recognition to compare the differences in global $x$ and $y$ coordinates on adjacent detector planes. 
%
%
The pattern recognition begins by filtering the clusters on all telescope planes, initially excluding the most upstream detector, to order them in a more efficient manner. This is done using the k-NN formalism, which builds an N-dimensional binary tree for each subsequent Timepix plane.
From this an arbitrary cluster is taken from the upstream Timepix plane and is used to seed the k-NN algorithm. The second telescope plane is then queried in order to return the nearest neighbours within a dynamic radial window $r_w$. This parameter has two contributions: a constant cluster search window $r_0$ and a term accounting for multiple scattering;

\begin{equation}
r_w (z, \alpha, \sigma_\theta) = r_0 + \Delta z~ {\rm tan} (\alpha \sigma_{\theta}) 
\label{eq:dynamicwindow}
\end{equation}

where $\Delta z$ is the displacement between two successive detector planes along the beam axis, $\sigma_\theta$ is the standard deviation of the the projected scattering angle in the $xz$ and $yz$ planes~\cite{cite:multscatter}
and $\alpha$ is a scaling factor, typically set to 3 to optimise the efficiency.  This window is dependent on $z$ and updates itself using the measured distances between planes. Each telescope plane consists of a 300\um thick silicon sensor, a Timepix chip (700\um thick) and a PCB with 100\um of copper and 1.3 mm of FR4. Adding up the contributions from these layers, the fraction of the radiation length seen by a particle traversing a telescope plane can be estimated as $x/X_{0} \simeq 2.6\%$.   The projected scattering angle is estimated using as input this radiation length and the track momentum (typically 180\gevc).  The multiple scattering contribution to the dynamic window is typically of the order of 10\um between the telescope arms for a 300\mm spacing, and will increase if a bulky DUT is mounted in this region, or if the distance between the two arms is increased.  In such a case it is possible to run the pattern recognition using a track extrapolation method, which projects the fitted track plane by plane, once the initial seed track with three clusters has been identified.  All clusters inside the window are stored as candidates and a fit is performed, only the candidate with the lowest $\chi^2$ is kept and added to the track. 

\subsection{Track Fit}

The track fit is performed using a least squares technique. As there is no magnetic field across the telescope, a track is represented in terms of a four-dimensional state vector $(x, t_x, y, t_y)$, where $(x, y)$ is the position at $z = 0$ and $t_x = d_x/d_z$, $t_y = d_y/d_z$ are the slopes with respect to the $z$ axis. After the full fit has been performed, the track probability is calculated from the $\chi^2$ per $n_{\rm dof}$ (number of degrees of freedom), where $n_{\rm dof} = 2n -4$, with $n$ being the number of points in the fit in $(x,y)$ coordinates and 4 representing the number of fitted parameters $(x,t_x,y,t_y)$.
A track probability cut of  $P_n(\chi^2) > 10^{-6}$ is applied, which removes about $20\%$ of tracks and results in very clean residual distributions. It should be noted that all the cuts in  Table~\ref{tab:params} are loose and the user can make tighter selections based on their requirements. 
The clustering, pattern recognition and trackfit are all configurable. Standard parameters which are used for the results presented in this paper are given in Table~\ref{tab:params}.


\begin{table}
\begin{center}
\begin{tabular}{l|c}
\hline
Parameter & Value \\
\hline
$\rm{ToT_{min}}$ [ADC counts] & 10 \\
$\rm{ToT_{max}}$ [ADC counts] & 880 \\
$r_o$ [$\mu$m] & 165 \\
$\alpha$ & 3 \\
min. clusters per track & 5 \\
Track $\chi^2$ Probability & $10^{-6}$ \\
\hline
\end{tabular}
\end{center}
\caption{Standard settings for configurable parameters used in clustering, tracking and track fitting}
\label{tab:params}
\end{table}

\section{Telescope Performance}

We characterise the performance of the telescope in terms of the rate of tracks and expected backgrounds available to the telescope user. For certain applications the resolution and/or time stamping may be considered to be of primary importance, and for this a loss in track reconstruction rate can be acceptable.  For other applications the primary importance may be the acquisition of high statistics, at the expense of a small degradation in the other telescope performance parameters. The expected track pointing resolution has been characterised as a function of track rate, and the time stamping facility of the telescope is described.

\subsection{Track Reconstruction rates and Backgrounds}
\begin{figure}
  \centering
  \includegraphics[width=0.5\textwidth]{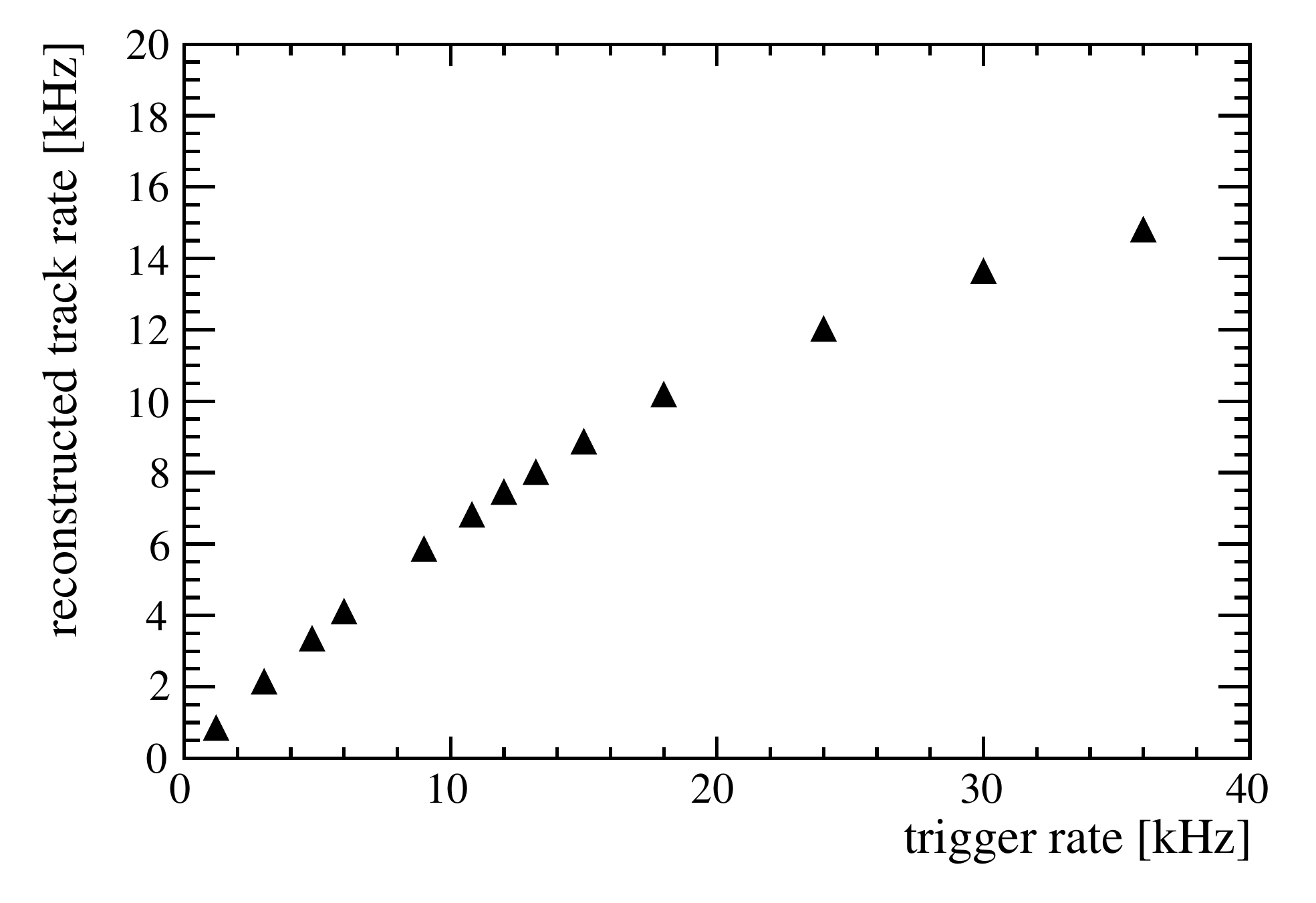}
  \caption{Rate of reconstructed tracks as function of the rate of scintillator triggers. Both track rate and trigger rate are to be understood as average rates during a spill (for a frame rate of 60\Hz).}
  \label{Fig:TrackRate}
\end{figure}

\begin{figure}
  \centering
  \includegraphics[width=0.5\textwidth]{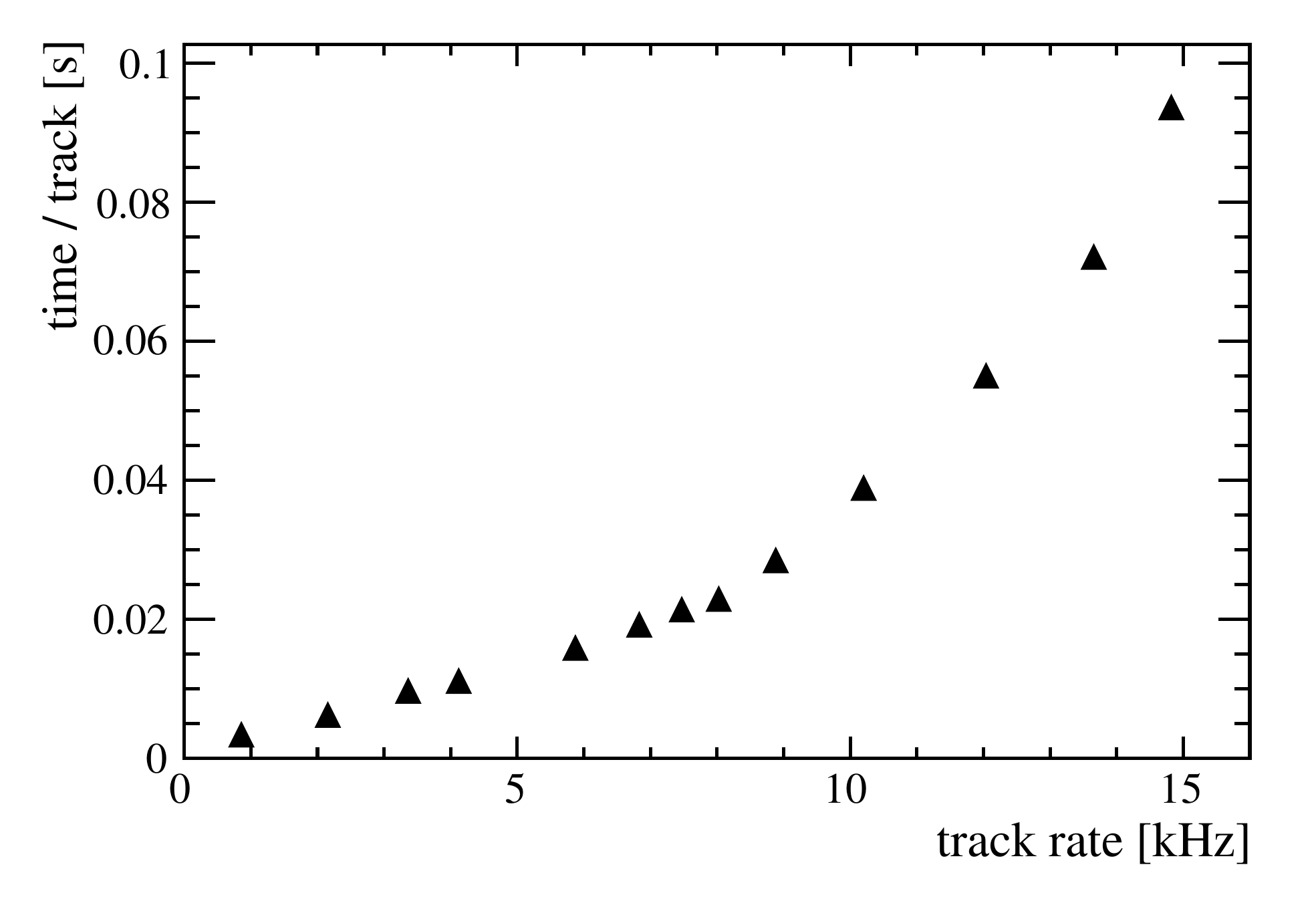}
  \caption{Average CPU time per reconstructed track as a function of reconstructed track rate as measured on an Intel Xeon single core (2.3\GHz).}
  \label{Fig:CpuTime}
\end{figure}

\begin{figure}
  \centering
  \includegraphics[width=0.5\textwidth]{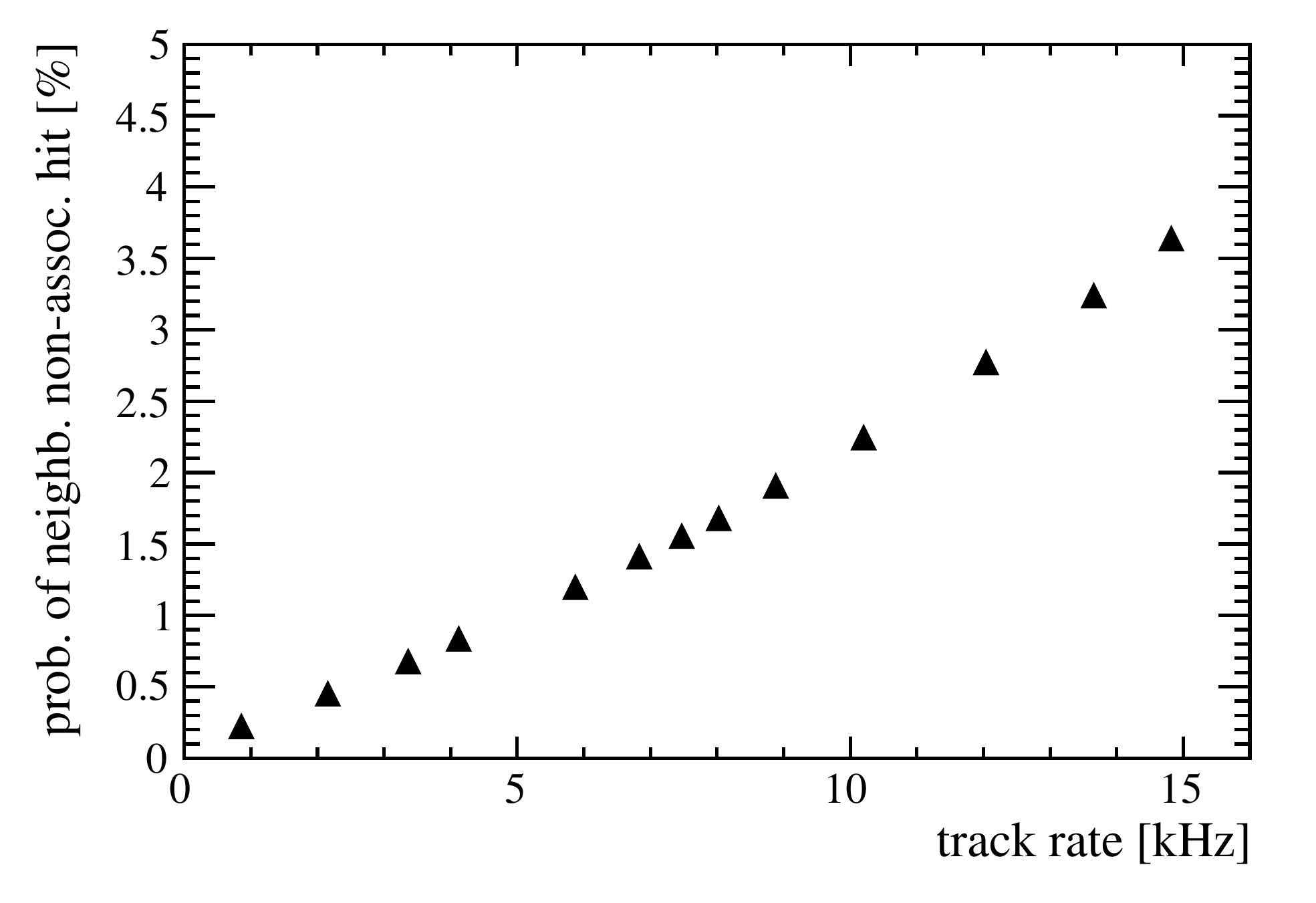}
  \caption{Probability of finding a non-associated cluster within a window of $\sim100\um$ within a track intercept as function of the rate of reconstructed tracks.}
  \label{Fig:FakeProbability}
\end{figure}

In Figures~\ref{Fig:TrackRate} to \ref{Fig:FakeProbability} some key telescope performance plots are presented.  In the standard mode of operation the shutter length is set by fixing the number of scintillator triggers.  Together with the 60 Hz readout rate this can be translated into an instantaneous track rate during the SPS spills, which last for about 10 seconds with a 40 second gap between spills. The number of reconstructed tracks available to the user does not match the scintillator rate due to the incomplete overlap of the planes and scintillators and imperfections in the pattern recognition. As the trigger rate increases to very high values the track reconstruction rate starts to level off and the CPU time per reconstructed track increases significantly at higher occupancy, as shown in Figure~\ref{Fig:CpuTime}.  This may be significant if the time taken to reconstruct the data offline is a limiting factor in the DUT analysis.  

When operating in the SPS beams  it has beenobserved that there are also additional clusters associated with the beam, coming from hadronic interactions in the material of the telescope and its support structures.  The rate of these clusters is at the level of about $15\%$ of the track induced clusters. This leads to the possibility that at the DUT there may be a neighbouring cluster close to the track intercept point in addition to the one generated by the track under consideration.  This effect may be of importance for certain studies (e.g. those wishing to measure very high efficiencies).  The size of the effect is shown in Figure~\ref{Fig:FakeProbability} which shows the chance that there may be a cluster within~100\um of the track intercept point which has been produced by a different process to that track.  The effect increases with occupancy, due to imperfections in the pattern recognition.   Depending on the requirements of the DUT and the analysis, a suitable working point can be selected.  The optimum working point of the telescope for typical conditions at the SPS H8 beamline gives a  reconstruction rate of about 7\kHz, leading to the accumulation of one million tracks at the device under test in about 12 minutes.

\subsection{Spatial Resolution}

\begin{figure}
\centering 
  \includegraphics[width=0.45\textwidth]{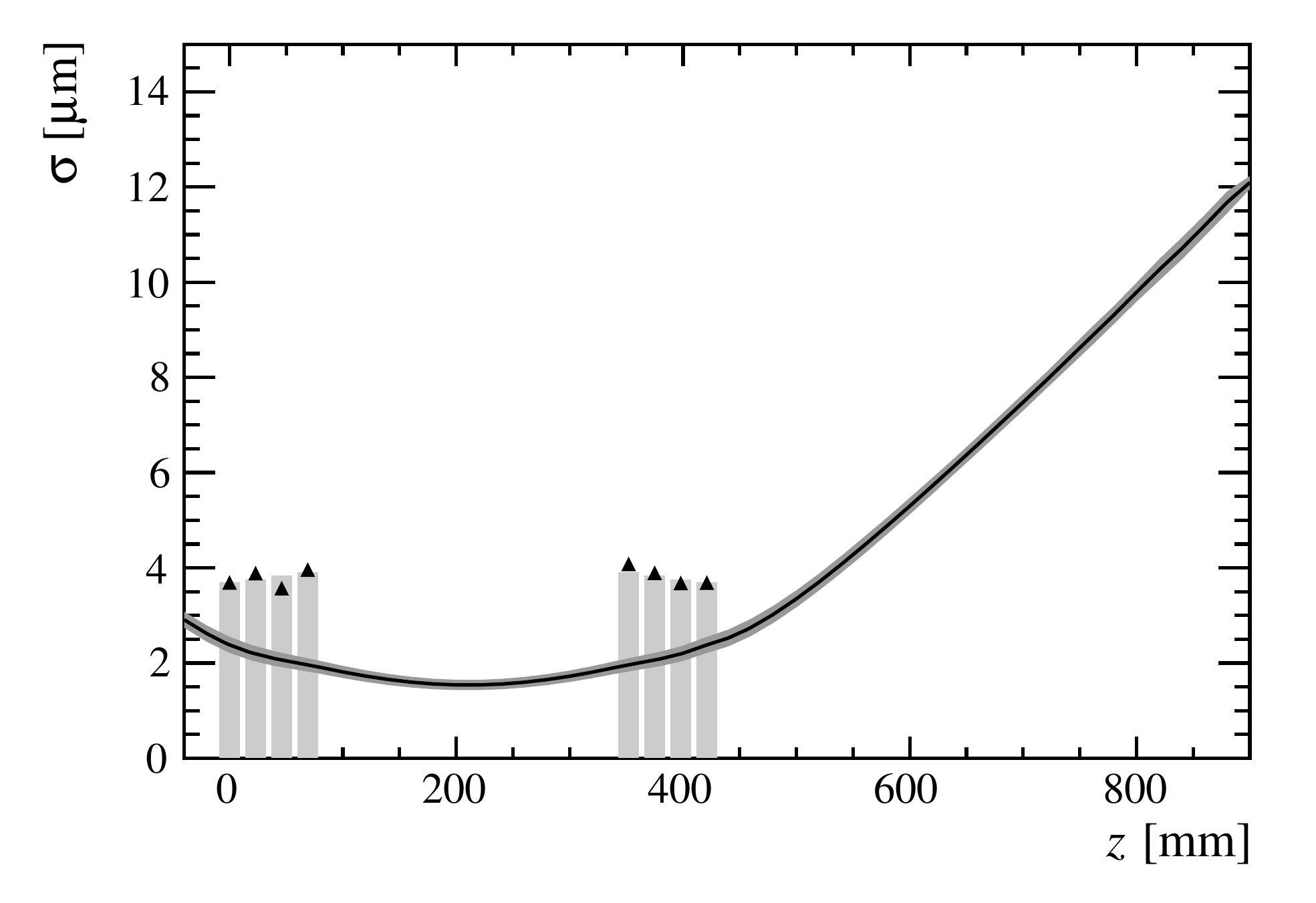}
  \includegraphics[width=0.45\textwidth]{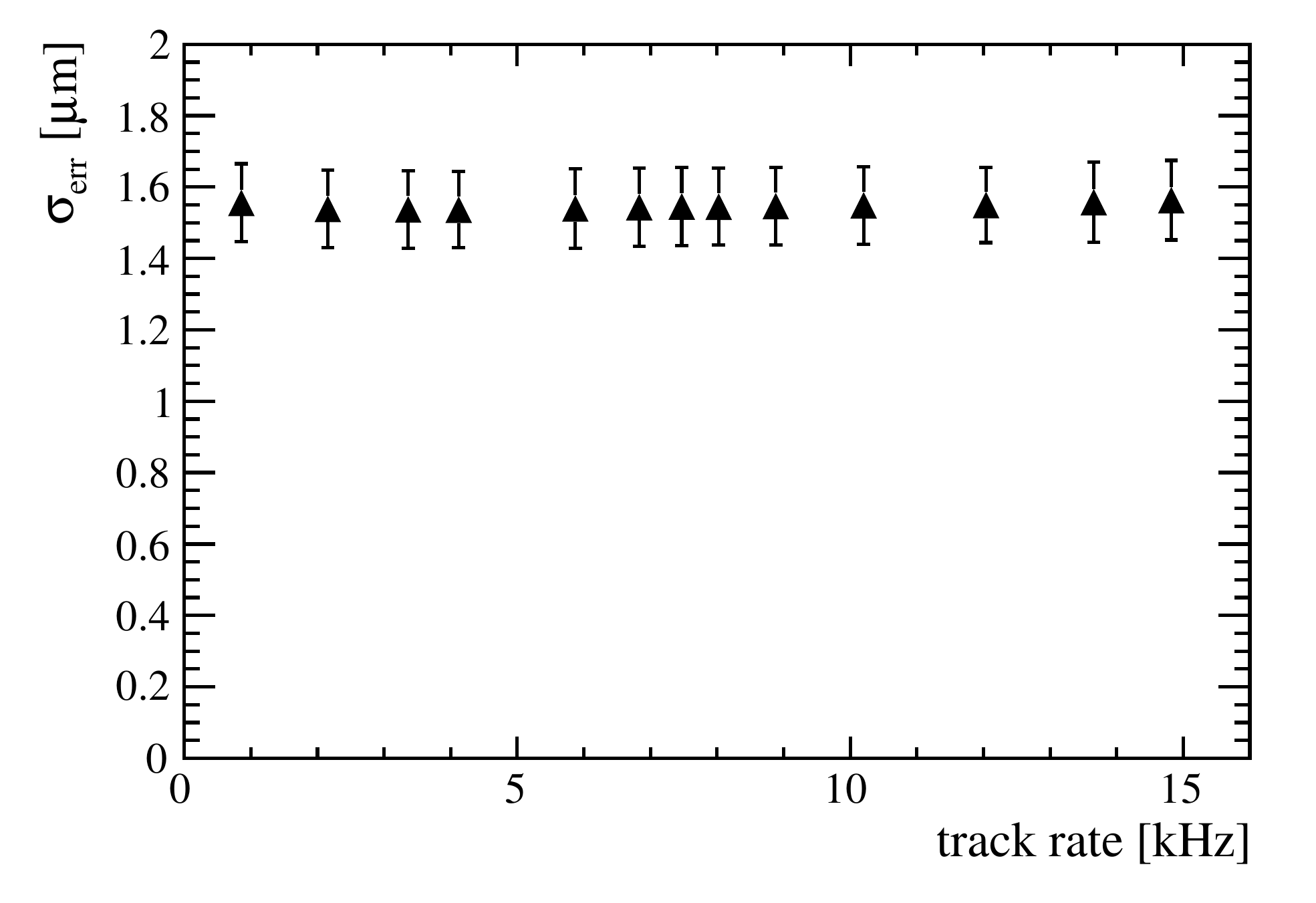}
  \caption{The left plot shows measured (triangles) and calculated (bars) biased residuals together with the predicted track pointing error (solid curve and dark-grey error band) as a function of the z position in the telescope.  The right plot the measured pointing error at the centre of the telescope as a function of the reconstructed track rate.}
  \label{Fig:ResolutionFit}
\end{figure}

The spatial resolution of the telescope depends on the single hit measurement error of a telescope plane, the material budget per plane \(x/X_{0}\), and the spacing between the individual planes. In order to estimate the track pointing error at the position of the device under test for a given configuration, the measured biased residuals in the telescope planes are compared to the results of a toy Monte Carlo simulation. In the simulation, the single hit measurement error is treated as a free parameter.  Its value is adjusted to minimize the difference between measured and calculated residuals. In order to determine \(x/X_{0}\), a dedicated experiment was performed in which the beam momentum was varied between 10 and 120\gevc. From this measurement we extract a material thickness of each plane of \(x/X_{0} = 0.028 \pm 0.1\) which is slightly higher than the value expected from the material composition.  At the default beam momentum (\(180\)\gevc) varying \(x/X_{0}\) between 2.6\% and 2.9\%  does not result in a significant variation of the extracted pointing error.  The distance between the planes within an arm is approximately 25\mm. In the configuration considered below, the two arms were separated by approximately 280\mm.
 
In Figure~\ref{Fig:ResolutionFit} biased residuals from data and simulation for a run with 50~triggers per frame are shown together with the predicted track pointing error.
The best pointing resolution, $\sigma_{\text{err}} = 1.54 \pm 0.11$\um, is achieved by a symmetric arrangement of the two telescope arms with the device under test at the centre. Due to small differences in thresholds between the planes, the residuals do not perfectly match the simulation; the level of the discrepancy is used for estimating the systematic error on $\sigma_{\text{err}}$.  The rightmost plot shows the measured pointing resolution at the centre of the telescope obtained as a function of the rate of reconstructed tracks.  For the chosen set of parameters (Table~\ref{tab:params}) no deterioration of the spatial resolution with increasing occupancy is observed.   These results are quoted for normal running conditions in the SPS with track momenta of 180\gevc.

The spatial resolution in the centre is fine enough to allow detailed investigation of sub-pixel sized structures in the device under test, such as the micromachined pores of a 3D sensor or the effect of active edge guardring processing.  Devices under test cannot only be placed in the centre of the telescope but also, at the cost of degraded resolution, behind the downstream arm. At 500\mm with respect to the centre of the telescope,  a resolution of approximately 8\um~is predicted.  

\begin{figure}
\centering 
  \includegraphics[width=1.0\textwidth]{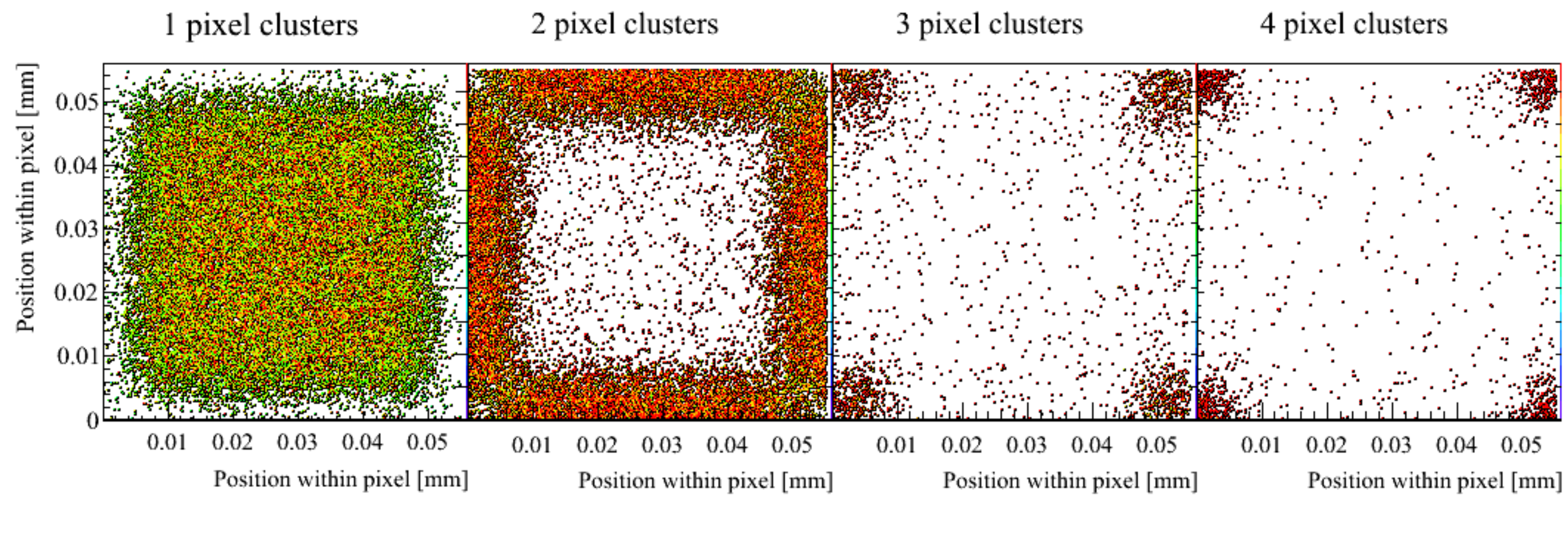}
  \caption{Track intercept position within the DUT pixel cell for 1,2,3 and 4 pixel clusters, weighted by the cluster ADC}
  \label{Fig:clusterpositions}
\end{figure}

Figure~\ref{Fig:clusterpositions} shows a typical example of how the telescope can be used to probe the behaviour within a pixel cell; in this case a Timepix bonded to a 150\um thick n-in-n sensor\footnote{The n-in-n sensor described was produced and boneded to the ASIC by VTT, Finland.} under study for the LHCb upgrade was placed at the centre of the telescope and the positions of the track intercepts corresponding to 1,2,3 and 4 pixel clusters, weighted by the ADC of the cluster is displayed.

\subsection{Track Time Stamping} 

\begin{figure}[h]
\centering
\includegraphics[width=13cm]{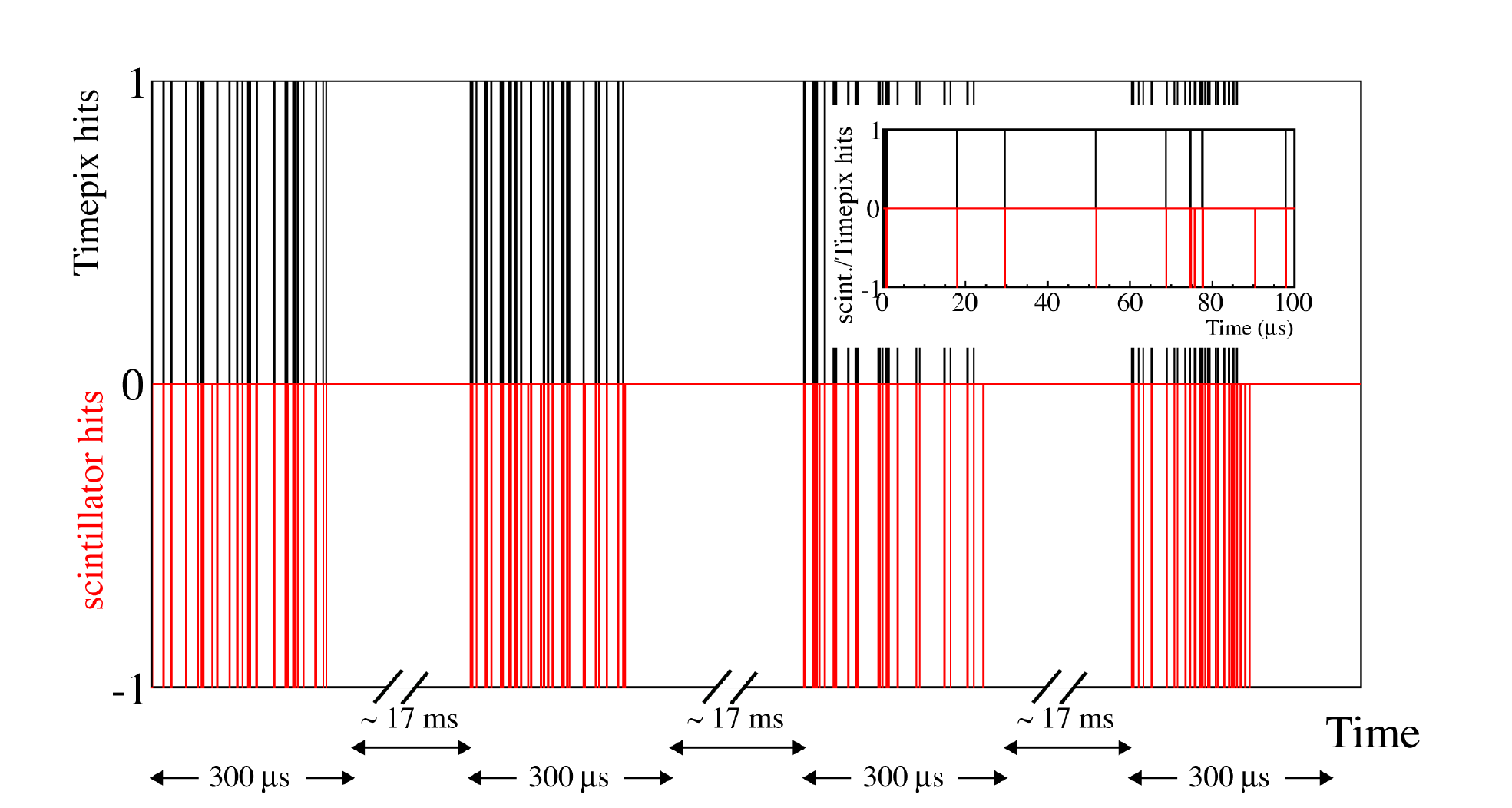}
\caption{The figure demonstrates the principle of the time alignment matching of hits in the scintillator system (black, towards the positive y axis) and Timepix timing plane (red, towards the negative y axis) for data from a typical run. The multiple data frames around which the Timepix shutter opens can be clearly seen.  The imperfect match is due to not all reconstructed tracks passing through the scintillator coincidence region, this can be seen more explicitly in the  insert where the time scale has been expanded.}
\label{fig:timing_example}
\end{figure}

Tracks in the telescope are time stamped using the most downstream Timepix plane in time of arrival (ToA) mode. This plane is mounted perpendicularly to the beam to minimise charge sharing and the associated increase in timewalk.  Due to the resulting inferior resolution this plane is not used for the tracking.  The ToA plane measures the number of clock cycles from the particle time of arrival until the shutter closing time, and is subject to a timewalk of typically 30\ns for normal running conditions.  This time stamp is refined using information from the scintillator and TDC system, by matching the track ToA to the nearest scintillator hit, as shown in Figure~\ref{fig:timing_example}.  This results in an improved track timestamping accuracy of about 1\ns.

The efficiency of the precise time stamping was measured, for 60 tracks per frame in the ToA plus scintillator fiducial region, and with a clock speed of 40\MHz, to be 98\%.   The main reason why the time stamping efficiency is not 100\% is because scintillator hits within 2.5 Timepix clock cycles of each other (at 40\MHz) are vetoed to ensure there is a unique candidate hit in the scintillator. This means the efficiency decreases at high beam intensities. With the SPS beam and a 40\MHz clock frequency, precise time stamps can still be obtained with efficiencies above 90\% for as many as 75 reconstructed tracks per frame.  Also, it should be noted that the time stamping can only be performed for the last 11810 counts in a frame, as tracks which pass through the plane in ToA mode before this time will saturate the Timepix counter.  Running the Timepix that is in ToA mode at a low clock frequency is appropriate for some applications where a long shutter time is desired. At low frequency the timewalk occupies fewer clock cycles, and at 10\MHz clock speed only scintillator hits within one clock cycle of each other need to be vetoed.   With a 10\MHz clock frequency and a slightly lower beam intensity it is possible to time stamp around 125 tracks per frame.

\section{Plans for Future Development}

The future direction of the Timepix Telescope is primarily driven by the development of the Timepix3\cite{timepix3} readout chip.  This is a new pixel readout ASIC being developed by the Medipix3 collaboration in the 130\nm CMOS process that has been successfully used to fabricate the current Medipix3 ASIC.  The new ASIC will continue to use the $256 \times 256$ 55\um square pixel footprint used by the current Timepix, Medipix2 and Medipix3 devices.   

A number of key performance elements, largely identified by the work on the Timepix Telescope, have been included.  Most critically all hits registered in the pixel matrix will contain both ToT and ToA information, allowing a continuing precise spatial resolution to be coupled to a significantly easier track recognition algorithm, removing the need for the additional timing plane and depreciating the importance of the scintillators.  Secondly the pixel front-end will operate much faster than the current device, increasing the accuracy of the ToA timing information by reducing timewalk and allowing the pixel matrix itself to provide timing resolution of 2\ns.  This will limit the role of the scintillator and TDC systems to providing triggers for the DUT readout system and a timing cross check to allow for a more robust integration and data analysis.  Furthermore the Timepix3 detector will operate in a data driven manner, generating data packets when track hits are registered up to a rate of 40\MHz across the chip.  This means there will be no significant readout dead time and the maximum sustained data rate will increase dramatically.  In practise it is expected that a Timepix3 system would be generally limited by the readout of the DUTs, as these will generally be test bench devices and not significantly optimised for readout speed.   It is envisaged that a standard readout developed for the detector will be incorporated into the Telescope system.

Beyond the development of the new ASIC, steps will be taken to improve the material thickness of the telescope, addressing the current limitation that the spatial resolution deteriorates for low energy beam lines.  An immediate saving in \(x/X_{0}\) can be achieved by redesigning the detector carrier boards to include no or minimal material behind the detector.  As this is required in any case as part of the Timepix3 development, it is relatively straightforward to ensure that the standard Timepix3 carrier PCBs have the material behind the hybrids optionally removed.  A more advanced strategy to reduce the material thickness will be to back thin the ASIC.  This has proved to be a successful but not trivial tactic in the past, with ASICs of this size being mechanically thinned from 700\um to 120\um in a process after the detector hybrids have been flip chip bonded.  This work with the Medipix footprint chips was pioneered by IEAP Prague.  It is expected that the reduction in material in the region of 420\um of silicon per plane will significantly improve the performance of the Telescope in lower energy facilities and make it a more generally useful instrument.  In addition the software package will be improved to incorporate full Kalman track fits working together with the k-NN algorithm, together with a Millipede~\cite{millipede} alignment package.

\section{Conclusions}

The Timepix Telescope described in this paper has been successfully operating at the CERN SPS North area in its current configuration throughout 2012.  For typical conditions during an SPS spill it provides up to 7.5\kHz of tracks with less than 2\um spatial and 1\ns temporal resolution at the centre of the telescope and more than 15\kHz of tracks if the time stamping requirement is dropped.  It has allowed the testing of a wide range of devices under study for future detector technologies.  The telescope will be extended in the future with the next generation of Timepix ASICs in order to dramatically improve the telescope readout rate and to improve the spatial resolution still further, in particular for operation with lower energy beamlines.  In this way the telescope will continue to support an international detector development R$\&$D programme.

\section*{Acknowledgements}

We gratefully acknowledge strong support from the CERN Medipix Group and many other members of the Medipix Collaboration, we also wish to thank the operators of the SPS beam line and North Area test facilities.  The research leading to these results has received partial funding from the European Commission under the FP7 Research Infrastructures project AIDA, grant agreement no. 262025.  We gratefully acknowledge the expert wire bonding support provided by Ian McGill of the CERN DSF bonding lab, and Igor Mandic and Vladimir Cindro for irradiating detector components at the Ljubljana irradiation facility.

\end{document}